\documentclass[aps, reprint, superscriptaddress]{revtex4-1}

\usepackage[pdftex,dvipsnames]{xcolor}

\usepackage[colorinlistoftodos,prependcaption]{todonotes}

\usepackage{bm}
\usepackage{graphicx}
\usepackage{amsmath}
\usepackage{mathbbol}
\usepackage{amssymb}
\usepackage[normalem]{ulem}
\usepackage[bookmarks,%
            linkbordercolor=white,%
	        urlbordercolor=white,%
	        colorlinks=true,%
	        linkcolor=black,%
            citecolor=black%
            ]{hyperref}
\usepackage{pdfpages}
\usepackage{pgffor}
\makeatletter
\AtBeginDocument{\let\LS@rot\@undefined}
\makeatother

\graphicspath{{./figures/}}

\begin{document}

\title{Proposal for a micromagnetic standard problem for materials with Dzyaloshinskii-Moriya interaction}

\author{David Cort\'{e}s-Ortu\~{n}o}
\email{d.cortes@soton.ac.uk}
\affiliation{Faculty of Engineering and the Environment, University of Southampton, Southampton SO17 1BJ, United Kingdom}
\author{Marijan Beg}
\affiliation{European XFEL GmbH, Holzkoppel 4, 22869 Schenefeld, Germany}
\author{Vanessa Nehruji}
\affiliation{Department of Physics, University of Durham, Durham DH1 3LE, United Kingdom}
\author{Leoni Breth}
\author{Ryan Pepper}
\affiliation{Faculty of Engineering and the Environment, University of Southampton, Southampton SO17 1BJ, United Kingdom}
\author{Thomas Kluyver}
\affiliation{Faculty of Engineering and the Environment, University of Southampton, Southampton SO17 1BJ, United Kingdom}
\author{Gary Downing}
\affiliation{Faculty of Engineering and the Environment, University of Southampton, Southampton SO17 1BJ, United Kingdom}
\author{Thorsten Hesjedal}
\affiliation{Department of Physics, University of Oxford, Oxford OX1 3PU, United Kingdom}
\author{Peter Hatton}
\author{Tom Lancaster}
\affiliation{Department of Physics, University of Durham, Durham DH1 3LE, United Kingdom}
\author{Riccardo Hertel}
\affiliation{Universit\'e de Strasbourg, CNRS, Institut de Physique et Chimie des Mat\'eriaux de Strasbourg, UMR 7504, F-67000 Strasbourg, France}
\author{Ondrej Hovorka}
\affiliation{Faculty of Engineering and the Environment, University of Southampton, Southampton SO17 1BJ, United Kingdom}
\author{Hans Fangohr}
\email{hans.fangohr@xfel.eu}
\affiliation{European XFEL GmbH, Holzkoppel 4, 22869 Schenefeld, Germany}
\affiliation{Faculty of Engineering and the Environment, University of Southampton, Southampton SO17 1BJ, United Kingdom}

\begin{abstract}

Understanding the role of the Dzyaloshinskii-Moriya interaction (DMI) for the
formation of helimagnetic order, as well as the emergence of skyrmions in
magnetic systems that lack inversion symmetry, has found increasing interest
due to the significant potential for novel spin based technologies.  Candidate
materials to host skyrmions include those belonging to the B20 group such as
FeGe, known for stabilising Bloch-like skyrmions, interfacial systems such as
cobalt multilayers or Pd/Fe bilayers on top of Ir(111), known for stabilising
N\'eel-like skyrmions, and, recently, alloys with a crystallographic symmetry
where anti-skyrmions are stabilised.  Micromagnetic simulations have become a
standard approach to aid the design and optimisation of spintronic and magnetic
nanodevices and are also applied to the modelling of device applications which
make use of skyrmions.  Several public domain micromagnetic simulation packages
such as OOMMF, MuMax3 and Fidimag already offer implementations of different
DMI terms. It is therefore highly desirable to propose a so-called
micromagnetic standard problem that would allow one to benchmark and test the
different software packages in a similar way as is done for ferromagnetic
materials without DMI. Here, we provide a sequence of well-defined and
increasingly complex computational problems for magnetic materials with DMI.
Our test problems include 1D, 2D and 3D domains, spin wave dynamics in the
presence of DMI, and validation of the analytical and numerical solutions
including uniform magnetisation, edge tilting, spin waves and skyrmion
formation. This set of problems can be used by developers and users of new
micromagnetic simulation codes for testing and validation and hence
establishing scientific credibility.

\end{abstract}

\maketitle

\section{Introduction}

In computational science so-called standard problems (or benchmark or test
problems) denote a class of problems that are defined in order to test the
capability of a newly developed software package to produce scientifically
trustworthy results. In the field of micromagnetism which, to a significant
extent, relies on results produced by computer simulations, the micromagnetic
modeling activity group ($\mathrm{\mu}$mag) at the National Institute of
Standards and Technology has helped to define and gather a series of such
standard problems for ferromagnetic materials on their website.~\cite{mumag}
Those five problems cover static as well as dynamic phenomena and over the
years more standard problems have been proposed including the physics of spin
transfer torque,~\cite{Najafi2009} spin waves~\cite{Venkat2013} and
ferromagnetic resonance.~\cite{Baker2017} Thus far, however, standard problems
for materials with Dzyaloshinskii-Moriya interaction (DMI) have not been
defined in the literature. Originally, the so-called DMI was phenomenologically
described by Dzyaloshinskii~\cite{Dzyaloshinskii1958,Dzyaloshinskii1964} to
explain the effect of weak ferromagnetism in antiferromagnets, and later it was
theoretically explained by Moriya~\cite{Moriya1960} as a spin orbit coupling
effect. The DMI effect is observable in magnetic materials with broken
inversion symmetry and can be present either in the crystallographic structure
of the material~\cite{Bogdanov1989,Leonov2015} or at the interface of a
ferromagnet with a heavy metal.~\cite{Crepieux1998,Fert2013,Wiesendanger2016}
In contrast to the favoured parallel alignment of neighbouring spins from the
ferromagnetic exchange interaction, the DMI favours the perpendicular alignment
of neighbouring spins.  The competition between these interactions allow the
observation of chiral magnetic configurations such as helices or skyrmions,
where spins have a fixed sense of rotation, which is known as chirality.

Skyrmions are localised and topologically non-trivial vortex-like magnetic
configurations.  Although they were theoretically predicted almost thirty years
ago,~\cite{Bogdanov1989} only recently have skyrmions started to attract
significant attention by the scientific community because of multiple recent
experimental observations of skyrmion phases in a variety of materials with
different DMI
mechanisms.~\cite{Muehlbauer2009,Yu2010,Milde2013,Romming2013,Leonov2016,Moreau-Luchaire2016,Pollard2017,Shibata2017,Nayak2017}
The magnetic profile of a skyrmion changes according to the kind of DMI present
in the material. Well-known skyrmionic textures are N\'eel skyrmions, Bloch
skyrmions and
anti-skyrmions.~\cite{Fert2013,Wiesendanger2016,Nayak2017,Hoffmann2017} The
former two are named according to the domain wall-like rotation sense of the
spins.

The fixed chirality of spins imposed by the DMI causes skyrmions to have
different properties from structures such as magnetic
bubbles~\cite{Nagaosa2013,Kiselev2011,Bernand-Mantel2017} or
vortices~\cite{Butenko2009,Mruczkiewicz2017,Chen2018}. In addition, the
antisymmetric nature of the DMI has an influence on the dynamics of excitations
such as spin waves, making them dependent on their propagation direction in the
material.

In this paper, we define a set of micromagnetic standard problems for systems
with different DMI mechanisms. This set of problems is aimed at verifying the
implementation of the DMI by comparing the numerical solutions from different
software with semi-analytical results from published
studies~\cite{Rohart2013,Bogdanov1994} where possible. In this context, we test
these problems using three open-source micromagnetic codes,
OOMMF,~\cite{OOMMFreprt} MuMax3~\cite{Vansteenkiste2014} and
Fidimag.~\cite{Fidimag} In Section~\ref{sec:analytic} we introduce our analysis
by defining the theoretical framework to describe ferromagnetic systems with
DMI, which is used to obtain numerical and analytical solutions.  Consequently
we describe the problems starting by the specification of a one dimensional
sample in Section~\ref{sec:1d-problem}, where the DMI has a distinctive
influence on the boundary conditions. Then in Section~\ref{sec:2d-problem} we
test the stabilisation of skyrmionic textures in a disk geometry for different
kinds of DMI. In Section~\ref{sec:sw-dynamics-problem} we compute the spin wave
spectrum of an interfacial system in a long stripe and show the antisymmetry
produced by the DMI. Finally, in Section~\ref{sec:3d-problem} we analyse a
skyrmion in a bulk system, where the propagation of the skyrmion configuration
across the thickness of the sample is known to be modulated towards the
surfaces because spins acquire an extra radial component.~\cite{Rybakov2013}


\section{The Dzyaloshinskii-Moriya interaction}
\label{sec:dmi}


The Dzyaloshinskii-Moriya
interaction~\cite{Dzyaloshinskii1958,Dzyaloshinskii1964,Moriya1960} (DMI) is a
spin orbit coupling effect that arises in crystals with a broken inversion
symmetry. In these materials the combination of the exchange and spin orbit
interactions between electrons leads to an effective interaction between
magnetic moments $\mathbf{S}_{i}$ of the form

\begin{equation}
H_{\mathrm{DM}} = \boldsymbol{D}\cdot\left( \boldsymbol{S}_{1}\times\boldsymbol{S}_{2}\right),
\label{eq:dmi-discrete}
\end{equation}

\noindent where the vector $\boldsymbol{D}$ depends on the induced orbital
moments. In general, the DM vector $\boldsymbol{D}$ will be non-zero, although
it is strongly constrained via Neumann's principle, that is, that the
Hamiltonian shares (at least) the symmetry of the underlying crystal system.
In this context, for two ions it is usually possible to strongly constrain the
direction of $\boldsymbol{D}$ through symmetry arguments, such as the ones
given in Refs.~\onlinecite{Yosida1996,Moriya1960}.  For example, if a mirror
plane runs perpendicular to the vector separating two ions, passing through its
midpoint, and if the separation is parallel to $z$, then this operation sends
$S_{1x} \leftrightarrow S_{2x}$, $S_{1y} \leftrightarrow S_{2y}$, and $S_{1z}
\leftrightarrow -S_{2z}$. This means that the transformation causes components
proportional only to $D_{z}$ to change signs, forcing $D_{z}$ to vanish but
leaving $D_{x}$ and $D_{y}$ nonzero.~\cite{Yosida1996}

When dealing with the continuum (micromagnetic) version of the DMI, the same
above considerations apply but may be generalized through the use of a
phenomenological approach based on Lifshitz invariants
(LIs)~\cite{Landau1980,Bogdanov2002}.  Systems featuring LIs range from
Chern-Simons terms in gauge field theories,~\cite{Lancaster2014} to chiral
liquid crystals,~\cite{Sparavigna2009} but also include magnetic systems
hosting DMIs. In this latter context they are said to describe inhomogeneous
DMIs~\cite{Bogdanov1989,Bogdanov2002} owing to the spatial variation of the
magnetization $\mathbf{m}$ that they describe, in the form

\begin{equation}
\mathcal{L}_{ij}^{(k)} =
m_{i}\frac{\partial m_{j}}{\partial k}
-
m_{j}\frac{\partial m_{i}}{\partial k},
\end{equation}

\noindent where $i,j,k\in\{x,y,z\}$. The precise forms of the LIs are dictated
by the crystal symmetry of the system and they determine the micromagnetic
expression of its DMI energy. In this continuum limit, the energy written in
terms of LIs encodes symmetry constraints elegantly using only a single
parameter $D$, by including the way in which the magnetisation (or spin)
changes along the different spatial directions. These continuum expressions of
the DMI energy are equivalent to the discrete version
(equation~\ref{eq:dmi-discrete}).

The DMI phenomenon also occurs at surfaces~\cite{Crepieux1998} and in
interfacial systems~\cite{Crepieux1998,Bogdanov2001,Fert2013} because of the
breaking of symmetries.  In the latter case, chiral interactions which lead to
LIs in the free energy of the system, could arise from broken symmetries
reflecting lattice mismatch, defects or interdiffusion between
layers~\cite{Bogdanov2001}.  A specific example of this~\cite{Fert2013} is
interfacial DMI arising from the indirect exchange within a triangle composed
of two spins and a non-magnetic atom with strong SO coupling.~\cite{Fert2013}
From the atomistic description of interfacial DMI it is possible to derive
expressions in the continuum based on LIs as shown in
Refs.~\onlinecite{Yang2015,Rohart2013}.

In general, for bulk and interfacial systems the application of the LI-based
continuum theory for inhomogeneous DMI, leads to the prediction of a rich
variety of non-collinear magnetic structures such as vortex
configurations.~\cite{Bogdanov1989,Bogdanov1994,Bogdanov2001,Leonov2015} An
extended discussion on the theory of the DMI and further examples are discussed
in Section~S1 of the Supplementary Material.

\section{Analytical model of chiral configurations}
\label{sec:analytic}

For the description of chiral structures in magnetic materials with DMI it is
customary to write the magnetisation in spherical coordinates that spatially
depend on cylindrical coordinates~\cite{Bogdanov1994,Leonov2015}

\begin{equation}
  \label{eq:m-cartesia-from-spherical-coordinates}
    \mathbf{m}=(m_x, m_y, m_z)=(\sin\Theta\cos\Psi,\sin\Theta\sin\Psi,\cos\Theta),
\end{equation}

\noindent where $(\Theta,\Psi)$ are spherical angles. In the general case,
$\Theta=\Theta(r,\phi,z)$ and $\Psi=\Psi(r,\phi,z)$ with $(r,\phi,z)$ being the
cylindrical coordinates. In the case of two dimensional systems or magnetic
configurations without modulation along the thickness of the sample, $\Theta$
and $\Psi$ are specified independent of the $z$-direction.

For a chiral ferromagnet without inversion symmetry, we are going to describe
stable magnetic configurations in confined geometries~\cite{Rohart2013} when
considering symmetric exchange and DMI interactions, an uniaxial anisotropy and
in some cases, an applied field. According to this, the energy of the magnetic
system is given by

\begin{align}
    E & = \int_{V} \text{d}V\, \bigg( A \sum_{i\in \{x,y,z\}} (\bm{\nabla} m_{i} )^{2}
                                      -K_{\text{u}} \left( \mathbf{m} \cdot \hat{z} \right)^{2} \\
      &  \phantom{= \int_{V} \text{d}V\,\bigg(} -\mu_{0} \mathbf{M}\cdot\mathbf{H}
                                                + w_{\text{DMI}}
                                \bigg), \nonumber
\label{eq:system-energy}
\end{align}

\noindent where $A$ is the exchange constant, $K_{\text{u}}$ is the uniaxial
anisotropy constant, $\mathbf{H}$ the applied field and the last term is the
DMI energy density, which can be written as a sum of Lifshitz
invariants~\cite{Bogdanov1989,Bogdanov1994,Leonov2015} (see
Section~\ref{sec:dmi}). For a material with symmetry class $T$ or $O$, the DMI
energy density is specified as

\begin{align}
    w_{\text{DMI}} & = D \left( \mathcal{L}_{zy}^{(x)} + \mathcal{L}_{xz}^{(y)} + \mathcal{L}_{yx}^{(z)} \right) \\
                   & = D \mathbf{m} \cdot \left( \bm{\nabla} \times \mathbf{m} \right).
\end{align}

\noindent For a thin film with interfacial DMI or a crystal with symmetry class
$C_{nv}$, located in the $x-y$ plane, the energy density of the DMI is

\begin{align}
    w_{\text{DMI}} & = D \left( \mathcal{L}_{xz}^{(x)} + \mathcal{L}_{yz}^{(y)} \right) \\
                   & = D \left( \mathbf{m} \cdot \bm{\nabla}m_{z} -
                              m_{z} \bm{\nabla}\cdot\mathbf{m}
                       \right).
    \label{eq:Cnv-DMI}
\end{align}

\noindent For a crystal with symmetry class $D_{2d}$, the DMI energy density
reads~\cite{Bogdanov1989}

\begin{align}
    w_{\text{DMI}} & = D \left( \mathcal{L}_{xz}^{(y)} + \mathcal{L}_{yz}^{(x)} \right) \\
                   & = D \mathbf{m} \cdot \left(
                              \frac{\partial \mathbf{m}}{\partial x} \times \hat{x}
                            - \frac{\partial \mathbf{m}}{\partial y} \times \hat{y}
                            \right).
\end{align}

Axially symmetric magnetic configurations that are uniform along the
$z$-direction can be found by substituting the magnetisation $\mathbf{m}$
(equation~\ref{eq:m-cartesia-from-spherical-coordinates}) into
equation~\ref{eq:system-energy}, with $\Theta=\Theta(r)$. Accepted solutions
for $\Psi$ are obtained according to the structure of the
DMI.~\cite{Bogdanov1989,Bogdanov1994,Leonov2015} For the $T$ class material,
$\Psi=\phi + \varphi$, with $\varphi=0,\pi$. In interfacial systems, the DMI
has the structure of a $C_{nv}$ symmetry class material, where
$\Psi=\phi+\varphi$ with $\varphi=\pm\pi/2$. For the $D_{2d}$ symmetry class an
accepted solution is $\Psi=-\phi+\pi/2$.


\begin{figure}[t!]
\includegraphics[width=\columnwidth]{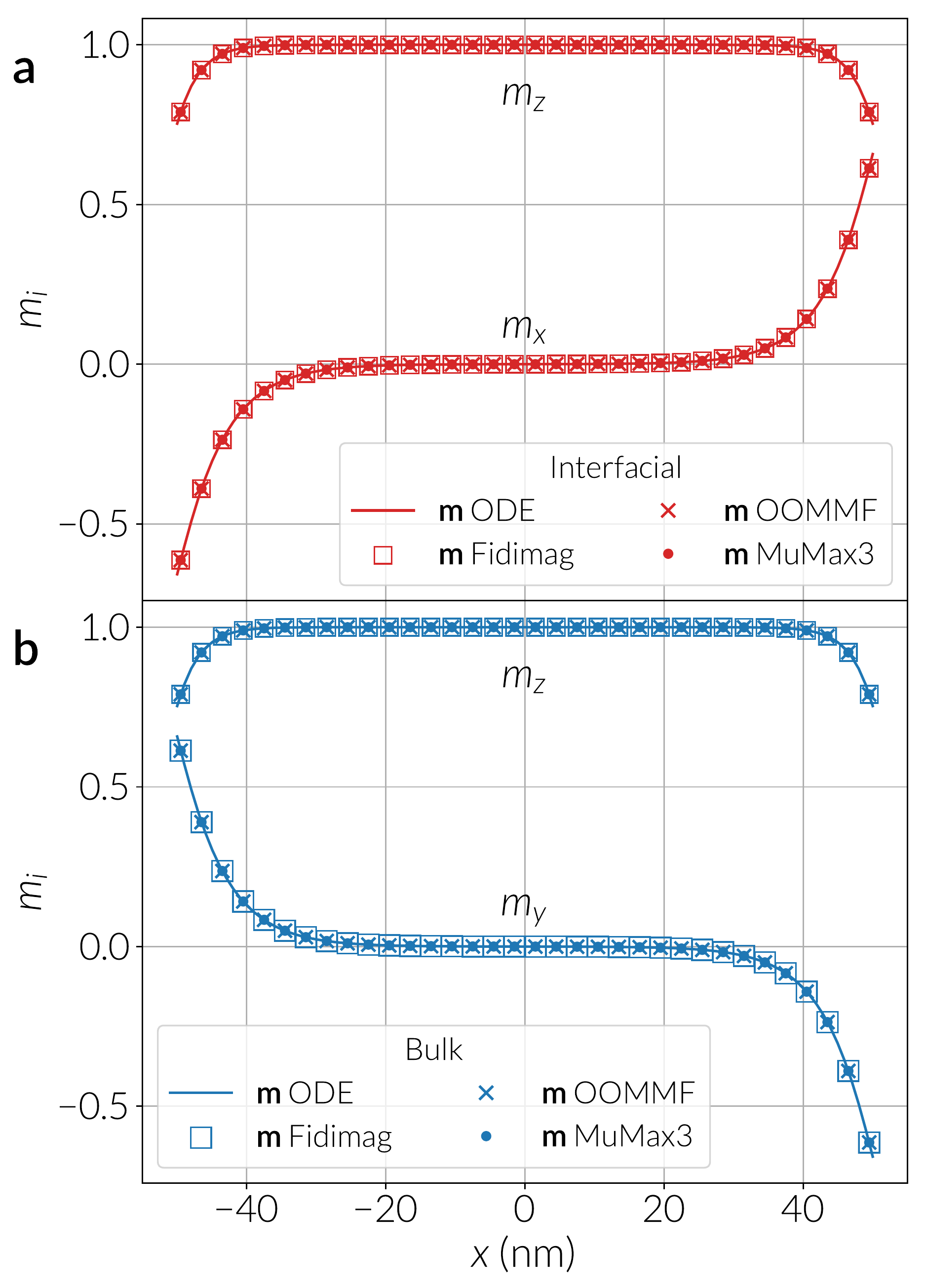}

\caption{Comparison of the magnetization components along a one-dimensional
permalloy-like wire with interfacial (a) or bulk (b) DMI. The wire long
axis is specified in the $x$-direction. The plot shows solutions from a
theoretical description of the system using an ordinary differential
equation (ODE), and solutions from the simulations using the OOMMF, Fidimag and
MuMax3 codes.}


\label{fig:1d-problem}
\end{figure}

\begin{table}[t!]
\begin{ruledtabular}
\begin{tabular}{ l r }
One-dimensional wire & \textrm{}\\
\colrule
Dimensions      & $100\:\text{nm}\times1\:\text{nm}\times1\:\text{nm}$\\
\colrule
Magnetic parameters & \\
\colrule
$A$             & $13\:\text{pJ m}^{-1}$    \\
$D$             & $3.0\:\text{mJ m}^{-2}$    \\
$M_{\text{s}}$  & $0.86\:\text{MA m}^{-1}$  \\
$K_{u}$         & $0.4\:\text{MJ m}^{-3}$
\end{tabular}
\end{ruledtabular}

\caption{\label{tab:problem-1d} One-dimensional problem specifications.}
\end{table}

\section{One-dimensional case: Edge tilting}
\label{sec:1d-problem}

In a one-dimensional magnetic system in the $x$-direction, where $x\in[0, L]$,
and at zero field, we can simplify the expression for the energy
(equation~\ref{eq:system-energy}) using $\Theta=\Theta(x)$ and $\phi=0$ (bulk),
where $\mathbf{m}=(\sin\Theta, 0, \cos\Theta)$, or $\phi=\pi/2$ (interfacial),
where $\mathbf{m}=(0, \sin\Theta, \cos\Theta)$. We therefore obtain the
following differential equation after minimising the energy with a variational
approach~\cite{Rohart2013}

\begin{align}
\frac{\text{d}^2 \Theta}{\text{d}x^2} & = \frac{\sin\Theta\cos\Theta}{\Delta^2} & \text{for } x(0) <x < x(L) \nonumber \\
\frac{\text{d}\Theta}{\text{d}x} & = \pm \frac{1}{\xi} & \text{at }x=x(0)\text{ or }x=x(L)
\label{eq:1d-diff-equation}
\end{align}

\noindent where $\Delta=\sqrt{A/K}$ and $\xi=2A/D$. The positive sign in the
boundary condition refers to the interfacial case and the negative sign to the
$T$ class material. We solve equations~\ref{eq:1d-diff-equation} using the
shooting method. For this, we refer to the alternative condition for $\Theta$
at $x=0$ or $x=L$ derived in Ref.~\onlinecite{Rohart2013}, which is valid when
the system has a large anisotropy and reads

\begin{equation}
    \sin\Theta=\pm\frac{\Delta}{\xi}.
\end{equation}

Depending on the chirality of the system, which can be observed from the
simulations, we fix the condition $\Theta(0)=\arcsin(\mp\Delta/\xi)$ and vary
$\text{d}\Theta(0)/\text{d}x$ until finding a solution that satisfies
$\Theta(L)=\arcsin(\pm\Delta/\xi)$. The upper sign~$+$ refers to the interfacial
case and the bottom sign~$-$ to the bulk DMI case.

In Fig.~\ref{fig:1d-problem}a and~b we compare results from the theory and
simulations of the one dimensional problem, for systems with interfacial
($C_{nv}$) and bulk ($T$) DMI, respectively.  For every case we use
permalloy-like parameters to test the problem, as specified in
Table~\ref{tab:problem-1d}. This material has associated an exchange length of
$L_{\text{ex}}=\sqrt{2A/(\mu_{0}M_{\text{s}}^2)}\approx5.3\:\text{nm}$ and a
helical length of $L_{D}=4\pi A/|D|\approx54.5\:\text{nm}$.  Simulations were
performed with the finite difference OOMMF, Fidimag and MuMax3 software. In our
examples we used a discretisation cell of $1\times1\times1\:\text{nm}^{3}$
volume, whose dimensions are well below the exchange length. The profile of the
$z$-component and either the $x$-component of the magnetisation, for the case
of interfacial DMI, or the $y$-component for the bulk DMI case, specify the
chirality of the magnetic configuration. To obtain the correct chirality in the
simulations, the DMI energy expression must be carefully discretised, as
explained in the Appendix~\ref{app:interfacial-dmi}.  For a one-dimensional
system, and since we are using common magnetic parameters, the major difference
between the $\mathbf{m}$ profiles of systems with different type of DMI, is the
orientation of the spin rotation. Therefore, for a crystal with $T$ or $D_{2d}$
symmetry, the profile of the $m_{y}$ component resemble the $m_{x}$ profile of
the interfacial DMI case, which is according to the spin rotation favoured in
the $T$ and $D_{2d}$ symmetries.  Accordingly, we only show the interfacial and
bulk DMI solutions in Fig.~\ref{fig:1d-problem}a and~b, respectively. In the
plots, data points from the simulations are compared with the solutions of
equations~\ref{eq:1d-diff-equation}. In general, OOMMF and Fidimag produce
similar results that perfectly agree with the theoretical curves obtained from
the solutions of the differential equations (as shown in
Fig.~\ref{fig:1d-problem}). Specifically, in the interfacial case the average
relative error (between the semi-analytical and simulation curves) for the
$m_{x}$ component is about 3.8\% and for $m_{z}$ is about 0.3\%. Equivalent
magnitudes are found for the bulk DMI system.  In the case of MuMax3, a similar
agreement  is found when imposing periodic boundary conditions along the
$y$-direction of the one-dimensional system because the DMI calculation is
implemented with Neumann boundary
conditions~\cite{Vansteenkiste2014,Leliaert2018} rather than free
boundaries~\footnote{This argument was discussed through an internal
communication with the MuMax3 team.  Free boundaries are going to be
available in future MuMax3 releases.} (see Section~S2 of the Supplementary
Material for a comparison when not using periodic boundaries).


\section{Two-dimensional case: magnetisation profile of a skyrmion}
\label{sec:2d-problem}


\begin{figure}[t!]
\includegraphics[width=\columnwidth]{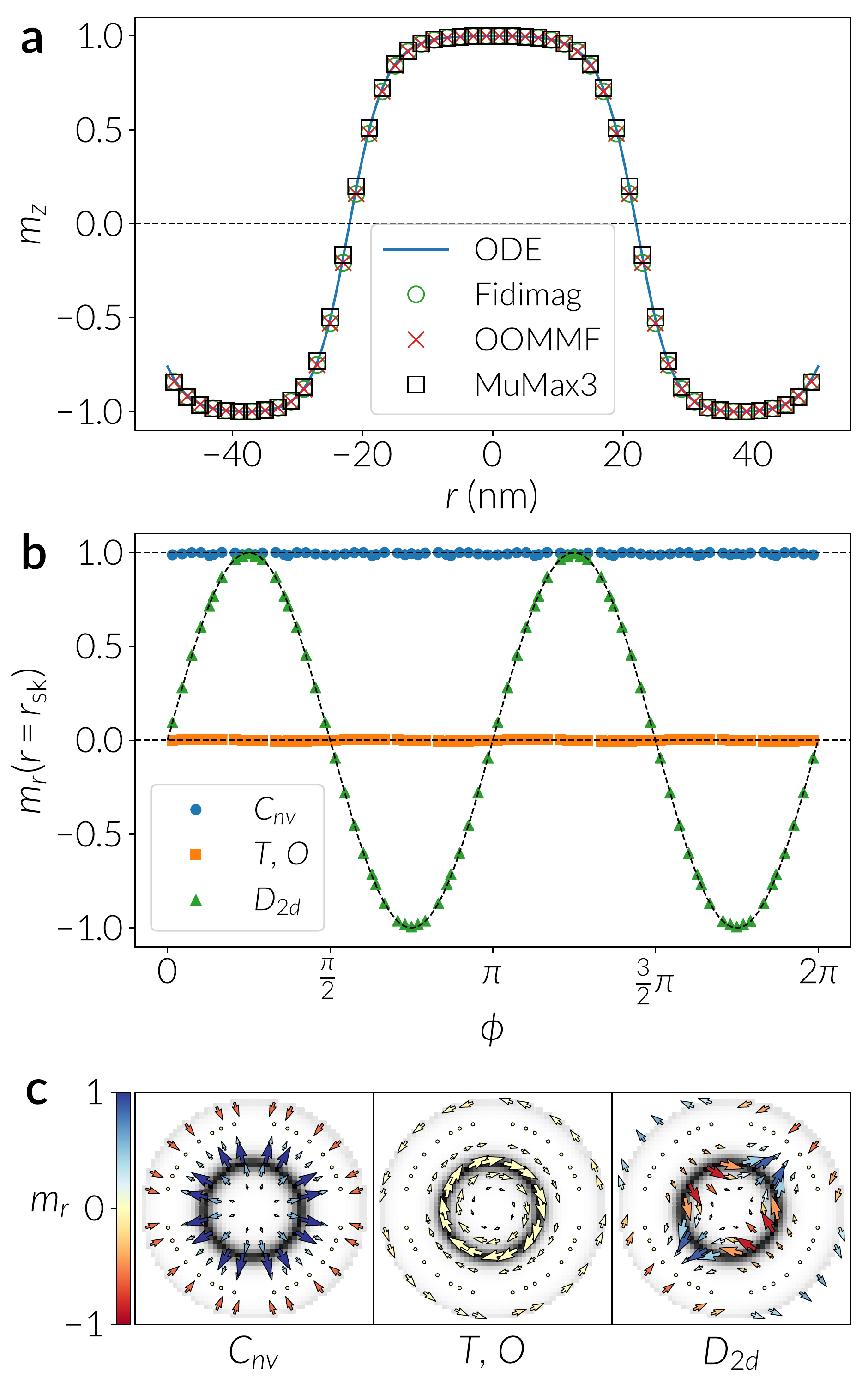}

\caption{Magnetization profile in permalloy-like disks with three different
DMIs from materials with symmetry class $C_{nv}$ (interfacial), $T$ or $O$
(bulk) and $D_{2d}$. (a) Comparison of the out-of-plane component of the
magnetization $m_{z}$ from the semi-analytical solution of the ordinary
differential equation (ODE) that describes the system, against results from
the simulations using three different codes.  (b) Radial component of the
magnetization in cylindrical coordinates, as a function of the azimuthal
angle, computed along the skyrmion radius $r_{\text{sk}}$ (where
$m_{z}=0$). The data points shown in the plot were obtained with the
Fidimag code and analytical solutions are drawn as thin dashed lines. (c)
Snapshots of the disk system for the three different DMI types. Arrows are
colored according to the radial component of the magnetization. The
background, in grey scale, illustrates the absolute value of the out of
plane component of the magnetization, where white means $|m_{z}|=1$ and black
means zero. The skyrmion centre is in the $+z$-direction, according to
plot~(a).}

\label{fig:2d-result}
\end{figure}

\begin{table}[t]
\begin{ruledtabular}
\begin{tabular}{ l r }
Two-dimensional disk & \textrm{}\\
\colrule
Radius          & $50\:\text{nm}$\\
Thickness       & $2\:\text{nm}$\\
\colrule
Magnetic parameters & \\
\colrule
$A$             & $13\:\text{pJ m}^{-1}$    \\
$D$             & $3.0\:\text{mJ m}^{-2}$    \\
$M_{\text{s}}$  & $0.86\:\text{MA m}^{-1}$  \\
$K_{u}$         & $0.4\:\text{MJ m}^{-3}$   \\
\end{tabular}
\end{ruledtabular}

\caption{\label{tab:problem-2d} Two-dimensional problem specifications.}
\end{table}

It has been shown in Refs.~\onlinecite{Carey2016,Beg2015,Rohart2013} that in a
confined geometry spins at the boundary of the system slightly tilt because of
the boundary condition and, due to the confinement, skyrmions can be stabilised
at zero magnetic field. Experimentally observed skyrmion configurations in
materials with three different types of DMIs have been reported in the
literature: (i)~Interfacial DMI, which favours N\'eel spin rotations and is
equivalent to the DMI found in systems with $C_{nv}$ crystal symmetry. (ii)~The
so called bulk DMI, which favours Bloch spin rotations and is found in
systems with symmetry class $T$ or $O$, such as FeGe. And, recently, (iii)~a
DMI found in systems with symmetry class $D_{2d}$ where structures known as
anti-skyrmions can be stabilised~\cite{Nayak2017,Hoffmann2017} (anti-skyrmions
have also been found in interfacial systems but they are best described within
a discrete spin formalism~\cite{Hoffmann2017,Camosi2017}). These three DMI
mechanisms can be described by a combination of Lifshitz invariants with a
single DMI constant.

We propose a two dimensional cylindrical system of $50\:\text{nm}$ radius and
$1\:\text{nm}$ thickness to test the stabilisation of skyrmions using the three
aforementioned DMIs, using permalloy-like magnetic parameters as in
Section~\ref{sec:1d-problem} (see Table~\ref{tab:problem-2d}).

We summarise in Fig.~\ref{fig:2d-result} results obtained for three different
skyrmion structures stabilised with the three kind of DMIs. These magnetic
configurations were simulated with finite difference codes, as shown in
Fig.~\ref{fig:2d-result}a, which shows a good agreement between them, thus we
only plot results from Fidimag in Fig.~\ref{fig:2d-result}b (see Section~S3 in
the Supplementary Material for a more detailed comparison). The three skyrmions
(see Fig.~\ref{fig:2d-result}c) are energetically and topologically equivalent,
and the magnetization profile depends only on the accepted solution for the
$\Psi$ angle when described in spherical coordinates. Therefore, the out of
plane component of the spins must match for the three configurations. Solving
this system analytically, we can calculate the $\Theta$ angle for the skyrmion
solution with the corresponding boundary condition~\cite{Rohart2013}, by
minimising equation~\ref{eq:system-energy}. We compare the out of plane
component of the spins, $m_{z}=\cos\Theta$, with that of the simulations by
extracting the data from the spins along the disk diameter, which we show in
Fig.~\ref{fig:2d-result}a.  As in the one-dimensional case, we observe the
characteristic canting of spins at the boundary of the sample.

To distinguish the three different systems, we compute the skyrmion radius
$r_{\text{sk}}$ by finding the value of $r$ where $m_{z}(r)=0$, and plot the
radial component of the spins $m_{r}$ (see
Appendix~\ref{app:cylindrical-comps}) located at a distance $r_{\text{sk}}$
from the disk centre. Since spins are in plane at $r=r_{\text{sk}}$, then
$\Theta=\pi/2$ and the radial component (see
Appendix~\ref{app:cylindrical-comps} and
equation~\ref{eq:m-cartesia-from-spherical-coordinates}) as a function of
$\phi$ is
$m_{r}(r_{\text{sk}},\phi)=\sin(\Theta(r_{\text{sk}}))\cos(\Psi-\phi)=\cos(\Psi-\phi)$.
Therefore, $m_{r_\mathrm{sk}}^{C_{nv}}=1$, $m_{r_\mathrm{sk}}^{T}=0$ and
$m_{r_\mathrm{sk}}^{D_{2d}}=\sin(2\phi)$. According to this, we see in
Fig.~\ref{fig:2d-result}b the simulated skyrmion radial profiles at
$r=r_{\text{sk}}$ for the $C_{nv}$, $T$ and $D_{2d}$ symmetry class materials,
which agree with the theory (shown in dashed lines and curves).

In Fig.~\ref{fig:2d-result}c we illustrate the three different configurations.
The radial component of the magnetisation is shown with a colormap and the
out-of-plane component is shown in grayscale, where white means $|m_{z}|=1$,
thus it is possible to distinguish the region that defines the skyrmion radius,
which is highlighted in black, and the slight spin canting at the disk
boundary.

For this two-dimensional problem, the three simulation packages produce similar
results and agree well in the solutions, matching the boundary conditions from
the theory. The theory predicts a skyrmion radius of
$r_{\text{sk}}\approx22.03\:\text{nm}$.  For the calculation of the skyrmion
radius in the simulation results, we use a third order spline interpolation of
the $m_{z}$ profile from Fig.~\ref{fig:2d-result}a. According to this, OOMMF
and Fidimag give a radius of $21.87\:\text{nm}$ and MuMax3 produces a radius of
22.1~nm, which is a slightly better approximation to the theoretical result.




\begin{figure}[t!]
\includegraphics[width=\columnwidth]{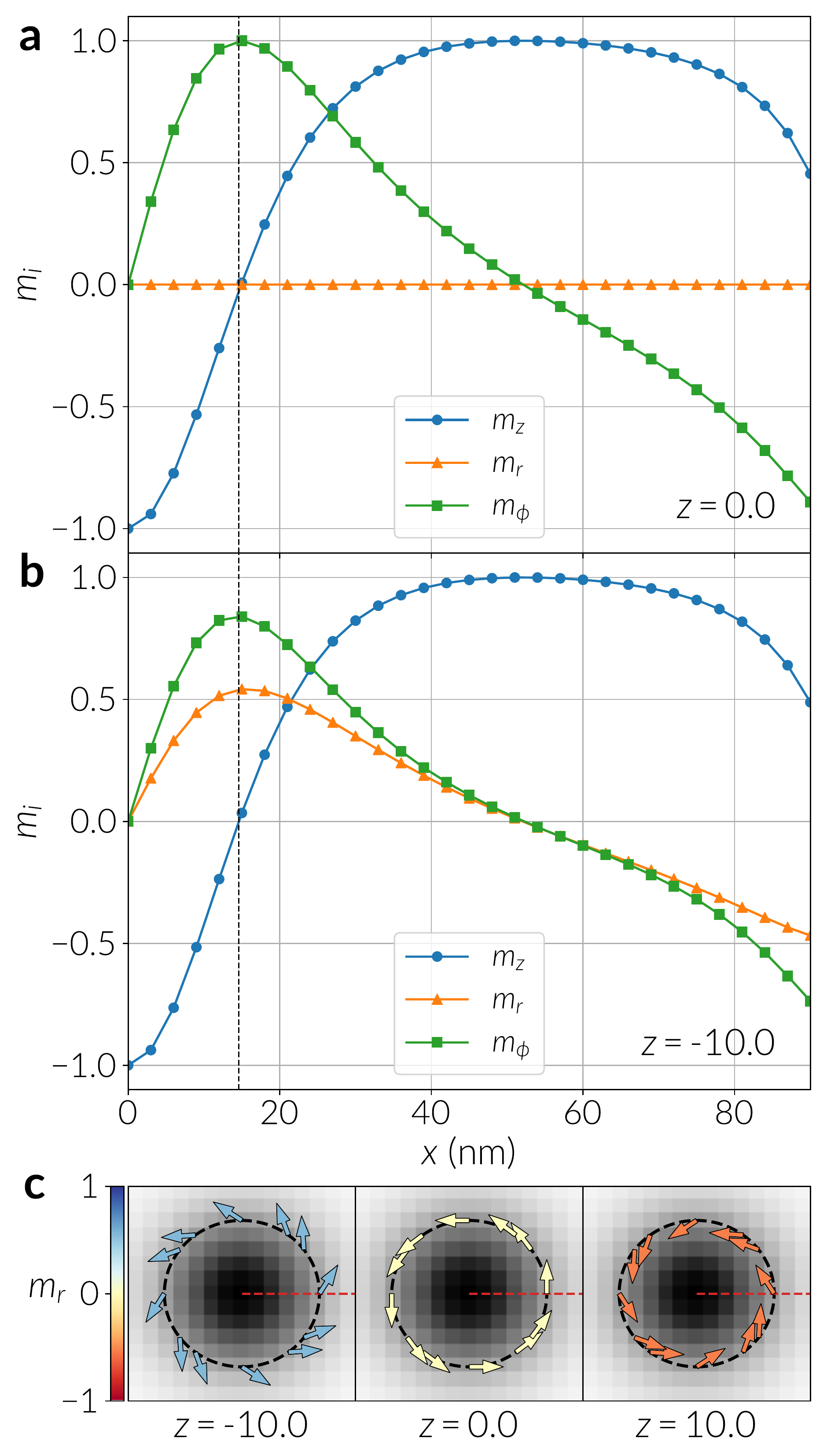}

\caption{Cylindrical components of the magnetization field of an isolated
skyrmion in a FeGe cuboid with periodic boundary conditions. These
numerical results were obtained using the Fidimag code. (a) Profiles across
the centre of an $x-y$ plane-cut of the cuboid (see red dashed line in
snapshots of plot~(c)) at $z=0\:\text{nm}$, which is the middle of the
sample across the thickness. (b) Profiles at the bottom surface of the
cuboid, which is the plane-cut at $z=-10\:\text{nm}$.  (c) Snapshots of the
magnetization profile at (from left to right) the bottom, middle and top
layers of the cuboid (which are plane-cuts) in the $z$-direction. The
sample is zoomed at the central region of the layers
($(x,y)\in[-22,22]\:\text{nm}\times[-22,22]\:\text{nm}$) where the skyrmion
centre is located. Spins are drawn in a circle defined by the skyrmion
radius, which is denoted by a dashed line, and are colored according to
their radial component. The background illustrates the out of plane
component of the magnetization $m_{z}$, where black means $m_{z}=1$.  }

\label{fig:result-3d}
\end{figure}

\begin{figure}[t!]
\includegraphics[width=\columnwidth]{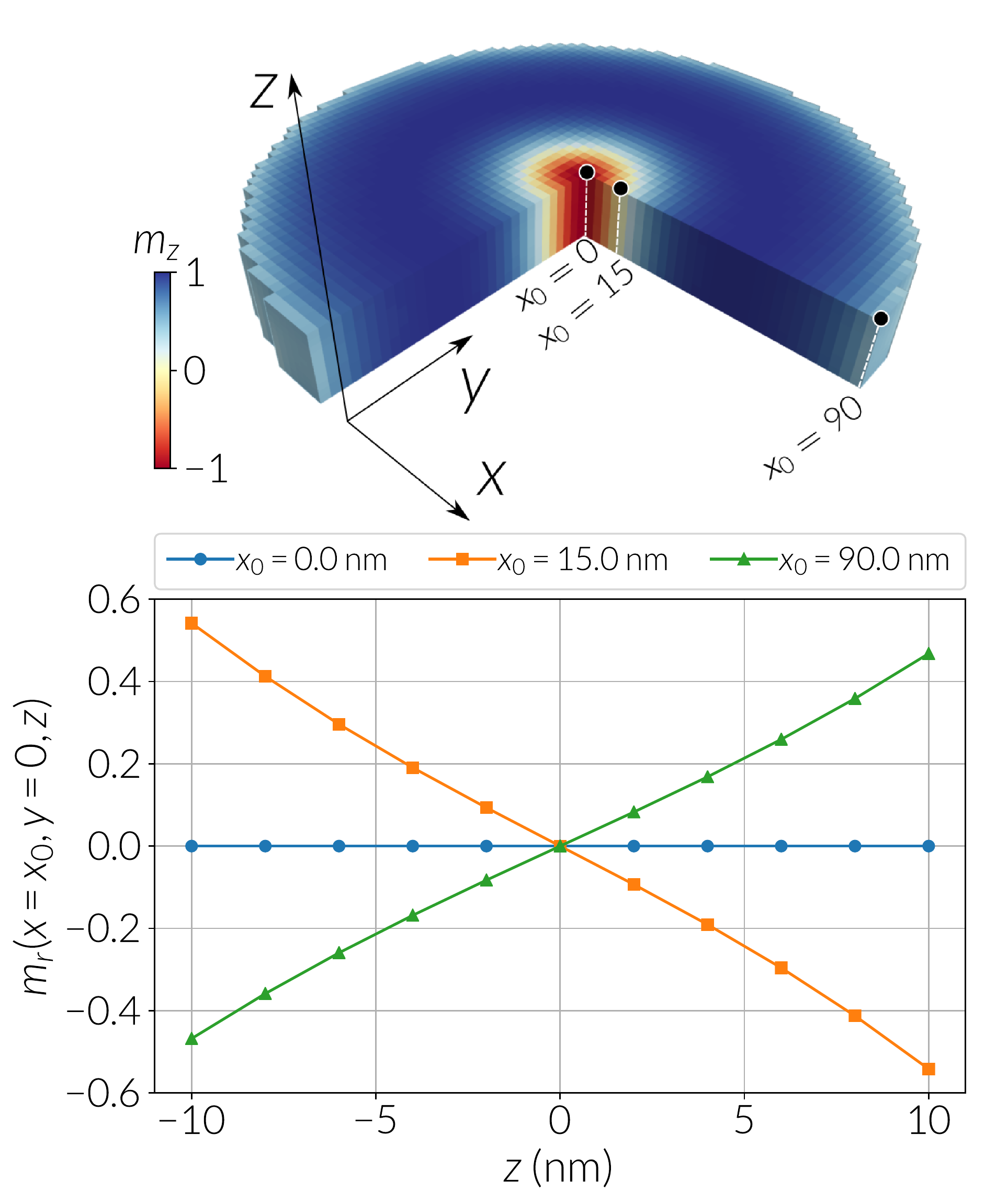}

\caption{Radial component of the magnetization across the cuboid thickness at
three different $(x_{0},y)$ positions for every plane cut in the
$z$-direction. The $y$-position is fixed at the centre of the system at
$y=0$.  The chosen $x$ coordinates are at the centre of the skyrmion
($x_{0}=0$), close to the skyrmion radius ($x_{0}=15\:\text{nm}$), which is
approximately 14.9~nm at the sample centre, and at the cuboid periodic
boundary ($x_{0}=90\:\text{nm}$). Data points were obtained from Fidimag
simulations. The top image shows the cylinder sample under study with the
three $(x_{0},y)$ positions marked as dots, and lines denoting where the
data is being extracted for every position.}

\label{fig:result-3d-linear-psi}
\end{figure}

\begin{table}[t]
\begin{ruledtabular}
\begin{tabular}{ l r }
Cylinder & \textrm{}\\
\colrule
Radius        & $90\:\text{nm}$ \\
Thickness     & $20\:\text{nm}$  \\
\colrule
Magnetic parameters & \\
\colrule
$A$                     & $8.78\:\text{pJ m}^{-1}$    \\
$D$                     & $1.58\:\text{mJ m}^{-2}$    \\
$M_{\text{s}}$          & $0.384\:\text{MA m}^{-1}$  \\
$\mu_{0}\mathbf{H}$     & $(0.0, 0.4, 0.0)\:\text{T}$
\end{tabular}
\end{ruledtabular}

\caption{\label{tab:problem-3d} Three-dimensional problem specifications.}
\end{table}

\section{Three-dimensional case: Skyrmion modulation along thickness}
\label{sec:3d-problem}

Skyrmions hosted in interfacial systems are in general effectively two
dimensional structures since these samples are a few monolayers thick. In
contrast, in bulk systems skyrmions can form long tubes by propagating their
double-twist modulation along their symmetry
axis,~\cite{Buhrandt2013,Milde2013,Rybakov2013,Carey2016,Charilaou2017} which
we will assume is along the sample thickness. Moreover, it has been shown by
Rybakov \textit{et al.}~\cite{Rybakov2013} that there is an extra spin
modulation along the skyrmion axis that can be approximately described by a
linear conical mode solution. This extra modulation is energetically favourable
in a range of applied magnetic field and sample thickness values, where the
latest is defined below the helix period $L_{D}$. This modulation is greatest
at the sample surfaces and is not present at the sample centre along the
thickness. It can be identified by an extra radial component acquired by the
spins, which is also maximal near the region where $m_{z}=0$ in every slice
normal to $z$ (see Fig.~\ref{fig:result-3d}c).

We define an isolated skyrmion in a FeGe cylinder of $180\:\text{nm}$ diameter
and $20\:\text{nm}$ thickness with its long axis (the thickness) in the
$z$-direction (see the system illustrated in
Fig.~\ref{fig:result-3d-linear-psi}).  We simulate the cylinder using finite
differences with cells of $3\:\text{nm}\times3\:\text{nm}\times2\:\text{nm}$
volume. Relaxing the system with an initial state resembling a Bloch
skyrmion,~\cite{Beg2015} and an applied magnetic field of
$B_{z}=0.4\:\text{T}$, we stabilise a skyrmion tube modulated along the
thickness of the sample. A characteristic parameter of FeGe is the helical
length $L_{D}=4\pi A/|D|\approx69.83\:\text{nm}$.

Results from Fidimag simulations are shown in Fig.~\ref{fig:result-3d}. We
obtain a skyrmion tube along the sample thickness with a radius that slightly
varies along the $z$-direction (we define as the radius where $m_{z}=0$ in a
$x-y$ plane-cut). In a slice of the cylinder at $z=0$, we compute a skyrmion
radius of $r_{\text{sk}}\approx14.9\:\text{nm}$, and this value decreases
towards the top and bottom surfaces down to $14.6\:\text{nm}$, which is
negligible in the scale of the chosen mesh discretisation. We emphasize that we
are not considering the demagnetizing field in this problem, which can enhance
this effect.

We analyse the magnetisation field profiles for different slices (different
$z$) by plotting the components of the spins located across a layer diameter
(or from the skyrmion centre), $(x,y)=(0, 0)\:\text{nm}$, up to the sample
boundary, $(x,y)=(90, 0)\:\text{nm}$, as shown by the red dashed line in every
snapshot of Fig.~\ref{fig:result-3d}c. To quantify the radial modulation of
spins, and because of the axial symmetry of skyrmions, we calculate the
cylindrical components $m_{r}$ and $m_{\phi}$ (see
Appendix~\ref{app:cylindrical-comps}).  Consistent with the results of
Ref.~\onlinecite{Rybakov2013}, Fig.~\ref{fig:result-3d}a reveals that at the
middle of the sample, i.e. at the $x-y$ plane-cut located at $z=0$, there is no
extra radial component of the spins, which is observed in a two dimensional
skyrmion. In addition, due the confined geometry the azimuthal component
slightly increases in magnitude towards the sample boundary with an opposite
sense of rotation than that of the skyrmion. Towards the sample surfaces,
located at $z=\pm10.0\:\text{nm}$, spins obtain an extra radial component that
increases linearly with the $z$ distance.  We illustrate this effect in
Fig.~\ref{fig:result-3d}b for the bottom layer at $z=-10.0\:\text{nm}$.
Interestingly, the maximum of $m_{r}$ and $m_{\phi}$ are slightly shifted with
respect to the $x$-position at $r_{\text{sk}}$, or where $m_{z}(x)=0$, which
can be seen from the data points around the dashed line of
Fig.~\ref{fig:result-3d}b. This same effect occurs at the top layer, but with
the radial component pointing inwards towards the skyrmion centre, thus $m_{r}$
looks like a mirror image of that of Fig.~\ref{fig:result-3d}b. Furthermore, we
notice that the radial component towards the boundary also changes sign as
$m_{\phi}$ does.  Snapshots with a zoomed view of the sample for the bottom,
middle and top layers are shown in Fig.~\ref{fig:result-3d}c. We show spins at
the skyrmion boundary, where $m_z=0$, colored according to their radial
component, and with the background colored according to the $m_z$ component.

The linear dependence of the radial component $m_{r}$ as a function of $z$
towards the surfaces is shown in Fig.~\ref{fig:result-3d-linear-psi}, where we
plot $m_{r}$ as a function of $z$ at three different $(x,y=0)$ positions in
every layer: the centre, $x=0\:\text{nm}$, close to the skyrmion radius
$r_{\text{sk}}$ (according to the discretisation of the mesh),
$x=15\:\text{nm}$, and at the sample boundary, $x=90\:\text{nm}$. These spatial
positions are shown as dots in the cylinder system at the top of
Fig.~\ref{fig:result-3d-linear-psi}, with lines denoting where the data is
being extracted. From the curves of Fig.~\ref{fig:result-3d-linear-psi} we
notice that the radial increment is maximal close to the skyrmion radius and is
slightly smaller, and with opposite orientation, at the cylinder boundary
normal to the radial direction.

Our results show that the skyrmion at the $z=0$ slice does not have a radial
modulation and the skyrmion size remains nearly constant across the sample
thickness.  Hence, it would be possible to use a two-dimensional model, similar
to the one used in Sections~\ref{sec:analytic} and~\ref{sec:2d-problem}, to
describe its profile. In performing this comparison (see Section~S4 in the
Supplementary Material) we noticed that the skyrmion in the cylinder system has
a larger skyrmion radius than the model predicts. In
Ref.~\onlinecite{Rybakov2013} an approximate solution is provided as an ansatz
for the $\Psi$ angle, which is based on a linear dependence on $z$.  Although
this approximation qualitatively describes the effects observed from the
simulations, a more accurate solution would be possible to obtain by taking the
general case $\Theta=\Theta(r,z)$ and $\Psi=\Psi(\phi,z)$, but it  generates a
non-trivial set of non-linear equations to be minimised. Because of the
consistent skyrmion size across $z$ it is likely that the dependence on $z$ in
the $\Theta$ angle only appears as a weak term or a constant, which
differentiates the solution from that of the two-dimensional model.

Testing this problem using the OOMMF code we obtain equivalent results with the
same skyrmion radius size. In the case of MuMax3, we had to use a slightly
different mesh discretisation, since the software only accepts an even number
of cells, which we adjusted to get a similar sample size. Results from MuMax3
simulations produce a skyrmion with larger radii compared to Fidimag and OOMMF,
with magnitudes of approximately 15.9~nm close to $z=0$ and 15.6~nm at the
cylinder caps. Nevertheless, the tendencies of the radial profiles of the
magnetisation are still close to the ones obtained with the other codes.
Details of these simulations are provided in Section~S5 of the Supplementary
Material. Although a cylinder system is also suitable for finite element code
simulations, a cuboid geometry is more natural to a finite difference
discretisation. Hence, we performed a similar study using a cuboid with
periodic boundary conditions. In general results on this geometry are
equivalent to the cylinder but with two main differences: the periodicity
removes the effects at the boundaries and the skyrmion is slightly larger in
radius. These solutions are shown in Section~S6 of the Supplementary Material.


\begin{figure}[t!]
\includegraphics[width=\columnwidth]{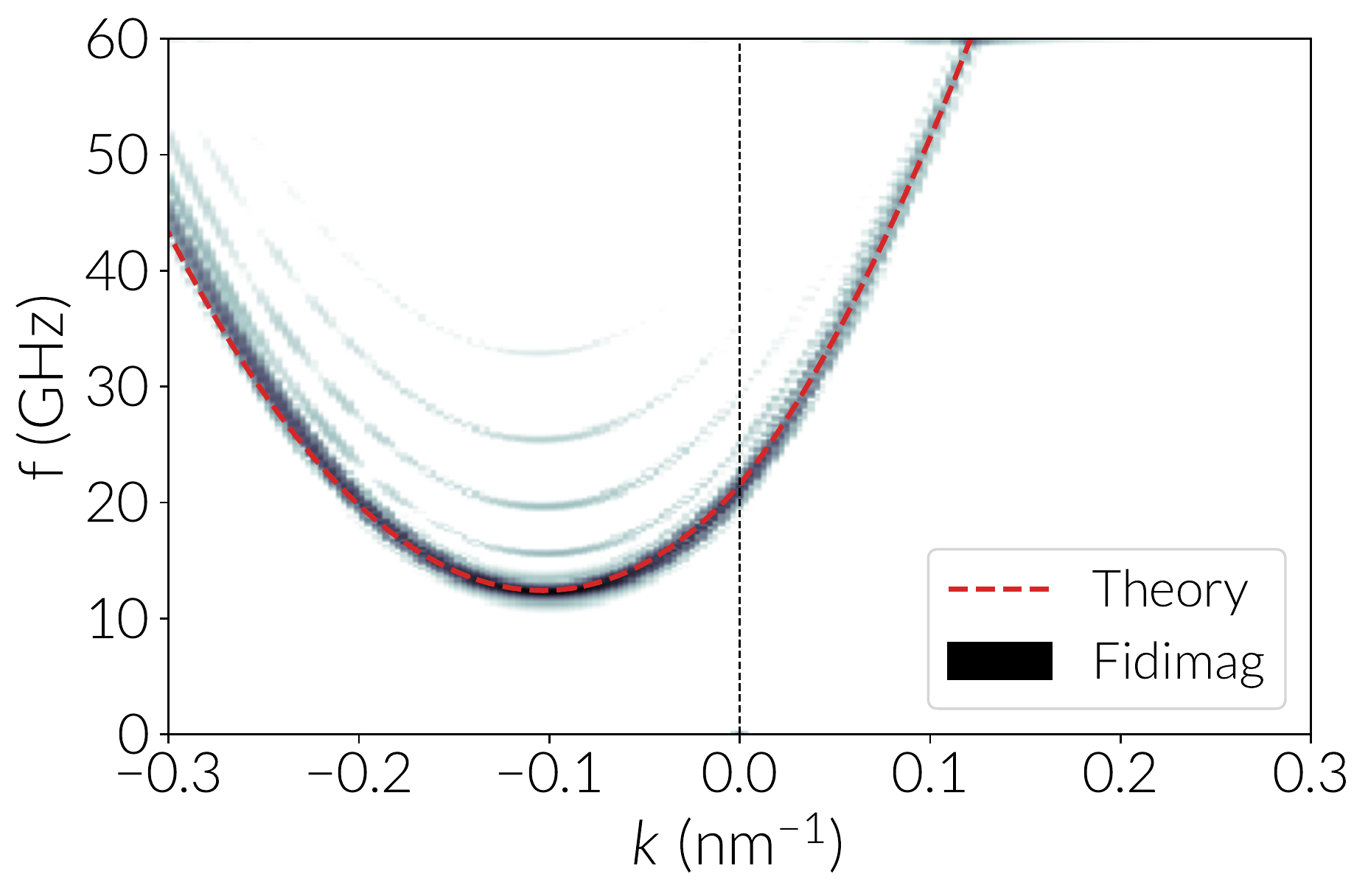}

\caption{Spectrum of Damon-Eshbach spin waves in a permalloy-like stripe
with interfacial DMI. The theoretical curve is computed from the theory
of Moon \textit{et al.}~\cite{Moon2013} and the intensity plot refers to the
result of the computer simulation of the system using the Fidimag code. The
intensity plot is given in logarithmic scale.}

\label{fig:spin-waves-problem}
\end{figure}

\section{Dynamics: asymmetric spin wave propagation in the presence of DMI}
\label{sec:sw-dynamics-problem}

Analyzing the dynamics of a magnetic system is a standard method to obtain
information about the magnetic properties of the material, such as damping or
the excitation modes of the system, among others. In particular, it is known
that the spin wave spectrum of a material with DMI saturated with an external
bias field, is antisymmetric along specific directions where spin waves are
propagating.~\cite{Cortes-Ortuno2013,Moon2013} These directions depend on the
nature of the DMI, and from the antisymmetry it is possible to quantify a
frequency shift from modes with the same wave vector magnitude but opposite
orientation, i.e. from waves travelling in opposite directions.  This frequency
shift depends linearly on the DMI magnitude of the material and hence it is a
straightforward method for measuring this magnetic parameter. This has been
proved in multiple experiments based on Brillouin light
scattering.~\cite{Belmeguenai2015,Di2015,Tacchi2017} Accordingly, a standard
problem based on spin waves offers the possibility to test the DMI influence on
the spin dynamics of the system.

\begin{table}[t!]
\begin{ruledtabular}
\begin{tabular}{ l r }
Two-dimensional stripe & \textrm{}\\
\colrule
Dimensions      & $2000\:\text{nm}\times200\:\text{nm}\times1\:\text{nm}$\\
\colrule
Magnetic parameters & \\
\colrule
$A$                                 & $13\:\text{pJ m}^{-1}$    \\
$D$                                 & $3.0\:\text{mJ m}^{-2}$    \\
$M_{\text{s}}$                      & $0.86\:\text{MA m}^{-1}$  \\
$\mu_{0}\mathbf{H}$                 & $(0.0, 0.4,0.0)\:\text{T}$   \\
$\mu_{0}\mathbf{h}_{\text{exc}}$    & $(0.04,0.0,0.0)\:\text{T}$   \\
$f_{0}$                             & $60\:\text{GHz}$   \\
$t_{0}$                             & $50\:\text{ps}$   \\
$\gamma$                            & $2.21\times10^{5}\:\text{Hz T}^{-1}$   \\
$\alpha$                            & $0.01$   \\
$\tau$                              & $4\:\text{ns}$   \\
$\Delta \tau$                       & $1\:\text{ps}$   \\
\end{tabular}
\end{ruledtabular}

\caption{\label{tab:problem-dynamics} Spin wave problem specifications.}
\end{table}

In interfacial systems spin waves propagating perpendicular to a saturating
bias field, which are known as Damon-Eshbach modes, exhibit antisymmetric
behaviour.~\cite{Cortes-Ortuno2013,Moon2013} To simulate this phenomenon, we
refer to the method specified in Ref.~\onlinecite{Venkat2013} and use
values from Table~\ref{tab:problem-dynamics},

\begin{enumerate}

    \item Define a thin stripe with the long axis in the $x$-direction and
        thickness along the $z$-direction.

    \item Saturate and relax the sample using a sufficiently strong bias
        magnetic field. If the relaxation is done with the LLG equation it is
        possible to remove the precessional term and use a large damping to
        accelerate the relaxation.

    \item Excite the system with a weak periodic field, based on a sinc
        function, in a small region at the centre of the stripe and applied in
        a specific direction $\hat{x}_{i}$,

        \begin{equation}
            \mathbf{h}_{\text{exc}} = h\, \text{sinc} \left( 2\pi f_{0} (t - t_{0}) \right)\hat{x}_{i}.
        \end{equation}

        We delay this signal by $t_{0}$, and then excite the system during
        the time $\tau$, saving the magnetization field every interval
        of duration $\Delta\tau$.

    \item Using the magnetization field files, we extract the dynamic
        components of the magnetization for a chain of magnetic moments along
        the $x$-direction across the middle of the stripe. The dynamic
        component is obtained by subtracting, from the excited spins, the
        components of the magnetic moments of the relaxed state obtained in
        step~2.

    \item We save these components in a matrix, where every column is a
        magnetization component, $m_x,m_y$ or $m_z$, of the spins across the spatial
        $x$-direction, and every row represents a saved time step saved in the
        previous step.

    \item Perform a two-dimensional spatial-temporal Fourier transform of the
        matrix, applying a Hanning windowing function~\cite{Kumar2012}.

\end{enumerate}

We define a permalloy stripe, with magnetic parameters specified in
Table~\ref{tab:problem-dynamics}, saturating the magnetization into the
$y$-direction. For this sample we take into account dipolar interactions. To
obtain a spectrum for positive and negative wave vectors, i.e., for waves
propagating in opposite directions, we excite the system in a small region of
$2\:\text{nm}$ width at the centre of the stripe, with a weak periodic signal
based on the cardinal sine wave function. We excite Damon-Eshbach spin waves by
applying the sinc field in the $x$-direction for a duration of $\tau = 4\:$ns.
According to Table~\ref{tab:problem-dynamics} we save $\tau / \Delta \tau =
4000$ steps to generate the spin wave spectrum.

The result of the spin waves simulation using Fidimag, after processing the
data, is shown in Fig.~\ref{fig:spin-waves-problem}. In the spectrum we compare
the result using the theory of Moon \textit{et al.}~\cite{Moon2013} for systems with
interfacial DMI. The asymmetry in the spin wave depends on the DMI sign.  To
compare the theoretical curve with the data from the simulations we calculated
the minimum in the dispersion relation for both curves. For the simulations we
calculated the peaks with largest intensity from the spectrum and fit the data
with a fourth order polynomial. The theory predicts that the minimum is located
at $k=-0.1036\:\text{nm}^{-1}$ with a frequency of $12.4098\:\text{GHz}$. From
the Fidimag simulation we estimate the minimum at $k=-0.1007\:\text{nm}^{-1}$
and $f=12.1690\:\text{GHz}$, which shows they are in good agreement.  As an
extra test we can notice from Ref.~\onlinecite{Cortes-Ortuno2013} that in
systems with crystallographic classes $T$ or $D_{2d}$, the Damon-Eshbach spin
waves will not be antisymmetric, however they would be for spin waves excited
along the field direction.

In Fig.~\ref{fig:spin-waves-problem} we observe the presence of extra modes
with a smaller intensity signal. These modes can be filtered by setting an
exponential damping towards the boundaries of the stripe~\cite{Venkat2018} to
avoid the reflection of spin waves. In the codes provided in the manuscript we
implemented functions for the damping with a simple exponential profile that
can be used to only obtain the main branch from the spectrum.

Using the OOMMF and MuMax3 codes for simulating spin waves in systems with DMI
produced equivalent results to Fig.~\ref{fig:spin-waves-problem}. With OOMMF
simulations we obtained a minimum at $k=-0.1001\:\text{nm}^{-1}$ and
$f=12.1796\:\text{GHz}$, while simulations performed with MuMax3 produce values
of $k=-0.1014\:\text{nm}^{-1}$ and $f=12.5232\:\text{GHz}$, which is a
slightly better approximation. Results for OOMMF and MuMax3 simulations, and
details about the numerical interpolation to the curves are shown in Section~S7
of the Supplementary Material.


\section{Conclusions}

We have proposed four standard problems to validate the implementation of
simulations of helimagnetic systems with DMI mechanisms found in crystals with
$C_{nv}$, $T$ and $D_{2d}$ symmetry class, where the former is also relevant in
interfacial systems. The strength of the three DMI types we use in the problems
can be quantified by a single DMI constant. For the one-dimensional and
two-dimensional problems we test the boundary condition in confined geometries,
which can be compared with analytical solutions. Moreover, profiles of
different skyrmionic textures, which vary according to the DMI kind, are
characterised by their radial profile, in particular at a distance $r$ from the
skyrmion centre where $m_{z}=0$, which we define as the skyrmion radius.
Further, in order to test the effect of the DMI on the dynamics of the systems,
we propose a problem based on the excitation of spin waves and the calculation
of their spectrum. In this case, we analyse Damon-Eshbach spin waves in a
stripe with interfacial DMI (or, equivalently, a crystal with $C_{nv}$
symmetry), which is known for being antisymmetric, and compare the solution
with analytical theory. Finally, we analyse an isolated skyrmion in a bulk
material with symmetry class $T$ in a cylinder. In this sample the skyrmion
profile propagates through the thickness and acquires an extra radial
modulation. We notice that this modulation is non-existent in a slice at the
middle of the sample along the thickness direction and increases linearly
towards the cylinder caps (normal to the $z$-direction). Additionally, it is
greatest at the skyrmion radius (where $m_{z}=0$), decreases to zero at the
skyrmion centre and towards the skyrmion boundary (in every slice), and is
present at the cylinder boundary (normal to the radial direction) with an
opposite orientation than the one within the skyrmion configuration.

Simulations in this study have been performed using codes based on the finite
difference numerical technique. Since many of the problems are compared with
semi-analytical calculations the results can be also applied to finite element
code simulations. Some finite element computations with our non-publicly
available software Finmag are shown in Section~S8 of the Supplementary
Material. In addition, we compared our data with the results from a non-public
finite-element code developed by R. Hertel, which is an entirely rewritten
successor of the TetraMag software~\cite{Hertel2007a,Kakay2010}. These results
are also shown in Section~S8, where we obtained an excellent quantitative
agreement.

With this set of problems we intend to cover the functionality of the DMI
interaction implemented in a micromagnetic code by testing boundary conditions,
energy minimisation, which can be achieved using LLG dynamics or minimisation
algorithms such as the conjugate gradient method, and spin dynamics. Overall,
the micromagnetic codes used in our testings significantly agree with expected
solutions and comparisons with the theory, thus our results substantiate
studies based on micromagnetic simulations with the three codes we have tested.
We hope this systematic analysis helps to promote the publication of codes in
simulation based studies for their corresponding validation and
reproducibility, and serve as a basis for more effective development of new
simulation software.

For the realisation of some of the problems, we have implemented new DMI
modules for MuMax3~\cite{Cortes2018} and
OOMMF~\cite{OOMMFa,OOMMFb,OOMMFc,Cortes2018} that take advantage of the
computer softwares framework, such as GPU implementation in MuMax3 or the
robustness of OOMMF. We have used the Jupyter OOMMF (JOOMMF) interface to drive
OOMMF and analyse data.~\cite{JOOMMF2017}. Scripts and notebooks to reproduce
the problems and data analysis from this paper can be found in
Ref.~\onlinecite{Cortes2018}.


\begin{acknowledgments}

This work was financially supported by the EPSRC Programme grant on
Skyrmionics~(EP/N032128/1), EPSRC’s Centre for Doctoral Training in Next
Generation Computational Modelling, http://ngcm.soton.ac.uk~(EP/L015382/1)
and EPSRC’s Doctoral Training Centre in Complex System
Simulation~(EP/G03690X/1), and the OpenDreamKit – Horizon 2020 European
Research Infrastructure project~(676541).

We acknowledge useful discussions with the MuMax3 code team.

\end{acknowledgments}

\appendix

\section{Finite difference discretisation for the DMI}
\label{app:interfacial-dmi}

Assuming a two dimensional film positioned in the $x-y$ plane, the energy
density $w$ for the interfacial DMI used in this study is modeled as

\begin{align}
w_{\text{DMI}} & = D \left( \mathcal{L}_{xz}^{(x)} + \mathcal{L}_{yz}^{(y)} \right) \\
               & = D \left( m_{x}\frac{\partial m_{z}}{\partial x} - m_{z}\frac{\partial m_{x}}{\partial x} +
                            m_{y}\frac{\partial m_{z}}{\partial y} - m_{z}\frac{\partial m_{y}}{\partial y}\right).
\label{eq:interfacial-dmi-lifshitz-invs}
\end{align}

The corresponding effective field of this interaction reads

\begin{align}
\mathbf{H}_{\text{DMI}} & = -\frac{1}{\mu_{0}M_{s}} \frac{\delta w_{\text{DMI}}}{\delta \mathbf{m}} \\
                        & = -\frac{2 D}{\mu_{0} M_{s}}
                             \left( \frac{\partial m_{z}}{\partial x}  \hat{x} +
                                    \frac{\partial m_{z}}{\partial y}  \hat{y} -
                                    \left[ \frac{\partial m_{x}}{\partial x} + \frac{\partial m_{y}}{\partial y}  \right] \hat{z}
                             \right).
\end{align}

When using finite differences, we can discretise the derivatives using a
central difference at every mesh site. Thus, for example, the central
difference for the first derivative of $m_{z}$ with respect to $x$ is

\begin{equation}
\frac{\partial m_{z}}{\partial x} \approx \frac{m_{z}(+x) - m_{z}(-x)}{2 \Delta x},
\end{equation}

\noindent where $m_{z}(\pm x)$ is the $m_{z}$ component of the closest magnetic
moment at the mesh site in the $\pm x$-direction and $\Delta x$ is the
mesh discretisation in the $x$-direction. We can then, for every mesh site,
collect all the terms related to the contribution to the field from its 4 mesh
neighbours in the plane where the system is defined. For instance, the field
contribution from the $+x$ mesh neighbour is

\begin{align}
    \mathbf{h}_{\text{DMI}}(+x) & = -\frac{2 D}{\mu_{0} M_{s}} \frac{1}{2 \Delta x}
                                     \left( m_{z}(+x)\hat{x} - m_{x}(+x)\hat{z} \right) \\
                                & = -\frac{2 D}{\mu_{0} M_{s}} \frac{1}{2 \Delta x}
                                     \left( \left[ \hat{z} \times \hat{x} \right] \times \mathbf{m}(+x) \right).
\end{align}

The contribution for the other neighbours have the same structure except the
denominator for the neighbours in the $y$-direction will have a factor of
$2\Delta y$ instead of $2\Delta x$ (similar for $z$). In addition, the cross
product is, in general, given by $(\hat{z}\times\hat{r}_{ij})\times\mathbf{m}$,
with $\hat{r}_{ij}$ the unit vector from the $i$ mesh site directed towards the
neighbour in the $j$-direction. Hence, the calculation for the field is similar
than that of the Heisenberg-like model, with an equivalent DMI vector of the
form $(\hat{z}\times\hat{r}_{ij})$.

It is important to mention that an interfacial DMI described within the
discrete spin model using the DMI vector
$\mathbf{D}_{ij}=(\hat{r}_{ij}\times\hat{z})$, in the continuum limit leads to
expression~\ref{eq:interfacial-dmi-lifshitz-invs}. However, when using finite
differences for the continuum description of the system, the calculation of the
field related to equation~\ref{eq:interfacial-dmi-lifshitz-invs}, which is
similar to the atomistic model calculation, has an opposite sign for the
equivalent DMI vector.

Similarly, for a $T$ class material, the finite differences discretisation
leads to a calculation of the micromagnetic DMI field using a vector
$\hat{r}_{ij}$. In the case of $D_{2d}$ symmetry, this vector is
$-\hat{r}_{ij}$ in the $x$-directions and $\hat{r}_{ij}$ in the $y$-directions.

\section{Cylindrical components}
\label{app:cylindrical-comps}

The cylindrical components of the magnetization are computed with a transformation
matrix according to

\begin{equation}
\begin{bmatrix}
   \cos\phi  & \sin\phi & 0 \\
   -\sin\phi & \cos\phi & 0 \\
   0            & 0           & 1 \\
\end{bmatrix}
\begin{bmatrix}
   m_x \\
   m_y \\
   m_z \\
\end{bmatrix}
=
\begin{bmatrix}
   m_r \\
   m_{\phi} \\
   m_z \\
\end{bmatrix}
\end{equation}

\noindent where $\phi=\arctan(y/x)$ is the azimuthal angle.

\bibliographystyle{naturemag_nourl}
\bibliography{abbreviated.bib}

\clearpage
\foreach \x in {1,...,19}
{%
\clearpage
\includepdf[pages={\x}]{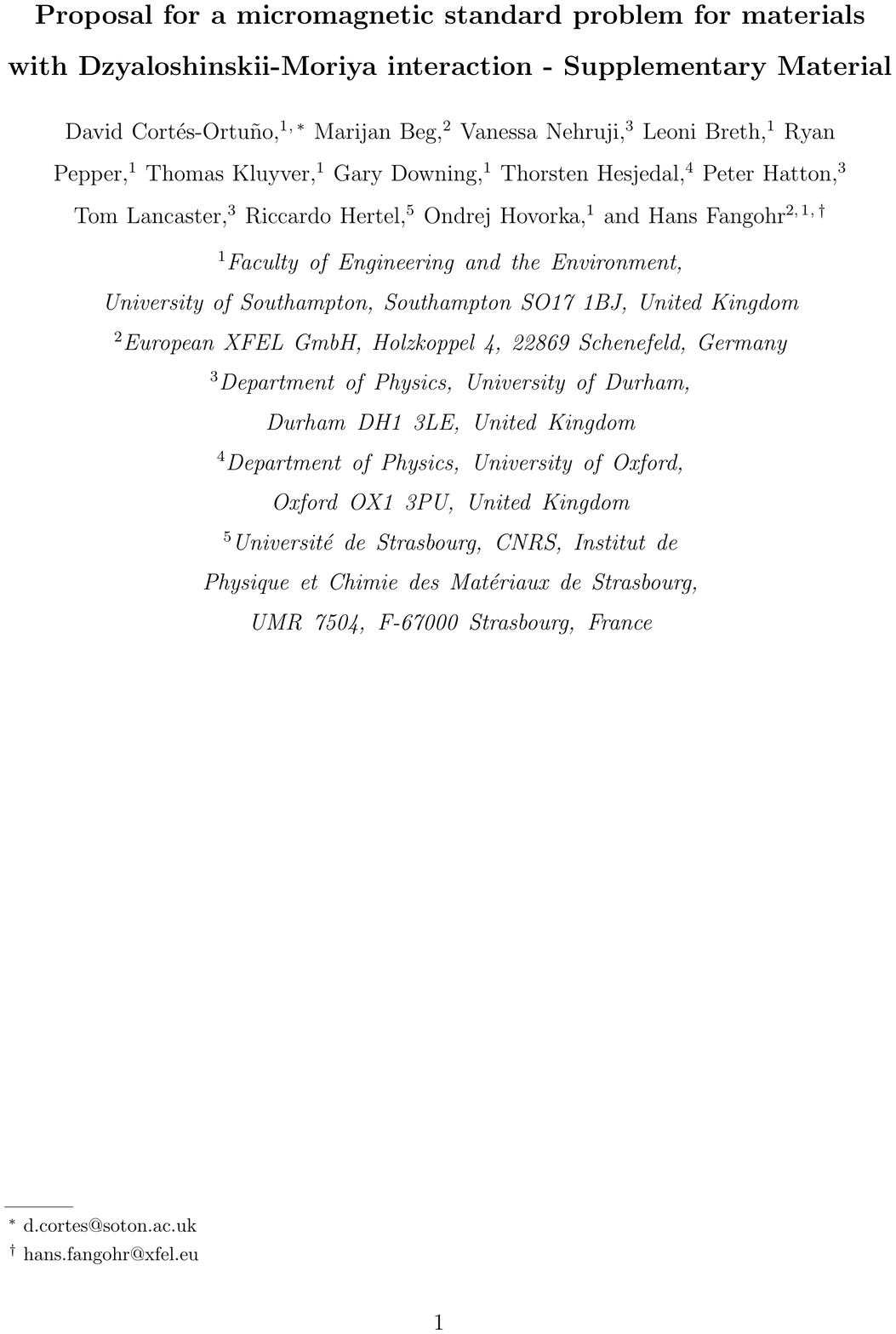} 
}
\end{document}